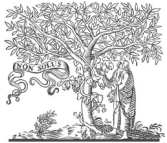

Contents lists available at ScienceDirect

# HardwareX

journal homepage: www.elsevier.com/locate/ohx

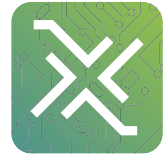

Hardware Article

# DIY hybrid SSVEP-P300 LED stimuli for BCI platform using EMOTIV EEG headset

Surej Mouli \*, Ramaswamy Palaniappan

*Data Science Research Group, School of Computing, University of Kent*

ABSTRACT

A fully customisable chip-on board (COB) LED design to evoke two brain responses simultaneously (steady state visual evoked potential (SSVEP) and transient evoked potential, P300) is discussed in this paper. Considering different possible modalities in brain-computer interfacing (BCI), SSVEP is widely accepted as it requires a lesser number of electroencephalogram (EEG) electrodes and minimal training time. The aim of this work was to produce a hybrid BCI hardware platform to evoke SSVEP and P300 precisely with reduced fatigue and improved classification performance. The system comprises of four independent radial green visual stimuli controlled individually by a 32-bit microcontroller platform to evoke SSVEP and four red LEDs flashing at random intervals to generate P300 events. The system can also record the P300 event timestamps that can be used in classification, to improve the accuracy and reliability. The hybrid stimulus was tested for real-time classification accuracy by controlling a LEGO robot to move in four directions.

© 2020 The Author(s). Published by Elsevier Ltd. This is an open access article under the CC BY license (http://creativecommons.org/licenses/by/4.0/).

Specification Table

| Hardware name | Hybrid BCI using SSVEP and P300 |
|---|---|
| Subject Area | Computer science, Brain-computing |
| Hardware Type | Electronic engineering, signal processing and computer science |
| Open Source License | CC BY 4.0 |
| Hardware Cost | < £250.00 |
| Source File Repository | https://osf.io/8bc5s/ |
| DOI | https://doi.org/10.17605/OSF.IO/8BC5S |

## 1. Hardware in context

Brain-computer interfaces (BCI) are widely used as an alternative communication mode to interact with the external world using brain signals [1–3]. There are various brain-activity measuring techniques such as functional near-infrared spectroscopy (fNIRS), electrocorticogram (ECoG), functional magnetic resonance imaging (fMRI) and electroencephalography (EEG). Amongst these techniques, EEG is widely used for BCI applications since the cost of the required hardware is low, non-invasive in nature and suitable for portable applications [4–6]. BCI applications are generally developed using EEG signals based on steady-state visual evoked potential (SSVEP) or transient evoked potential (such as P300) as these give higher recognition rates [7–8]. SSVEP is a natural response to a visual stimulus flickering at a constant frequency and produces brain signals with the same frequency as that of the stimulus in the visual cortex. As for P300, it is a transient event-related potential, usually generated by the oddball paradigm (target and non-target stimuli are shown with target stimuli having lower probability of occurrence). This component occurs as a positive deflection in EEG approximately 300 ms after the target object has been perceived by the user.

---

\* Corresponding author.

Majority of the BCI platforms use single EEG paradigm, which may not work for all users and could also generate false recognition. In the recent years, BCI systems have been enhanced by using multiple paradigms to improve accuracy and speed to control external applications [8–12]. Possibility of detecting both P300 component and SSVEP activity simultaneously has been confirmed by previous studies [13]. Majority of the hardware for P300 and SSVEP stimuli are based on computer screens that are limited to screen refresh rates, which also reduce portability and induce visual fatigue [8,14–15]. In this study, we have designed a fully portable hybrid system to evoke both P300 and SSVEP with a high-precision dedicated hardware platform. In this hybrid hardware, SSVEP was used as the primary response and P300 as a corrective mechanism in classification. Four individual chip-on-board (COB) green LED radial stimuli were used in this study for SSVEP elicitation, which was individually controlled by microcontroller platforms to generate precise flicker frequencies.

The stimulus platform was controlled by independent control systems as shown in Fig. 1.1 Four radial green stimuli for eliciting SSVEP were controlled by four Teensy 32-bit microcontroller module and four red stimuli for P300 were controlled by a separate Teensy module. Teensy also sent the flicker event markers for each flash separately to the recording software.

## 2. Hardware description

For multiple SSVEP elicitation, four independent Teensy microcontroller platforms were used to generate the flicker frequencies 7, 8, 9 and 10 Hz using green radial stimulus. Inside each radial ring, high-power red LED was placed to evoke P300 events, which were event marked along with the SSVEP EEG data. The red flashes were presented with random timings controlled by an individual Teensy module. The flash events were transferred as serial data from the microcontroller to the EEG recording software.

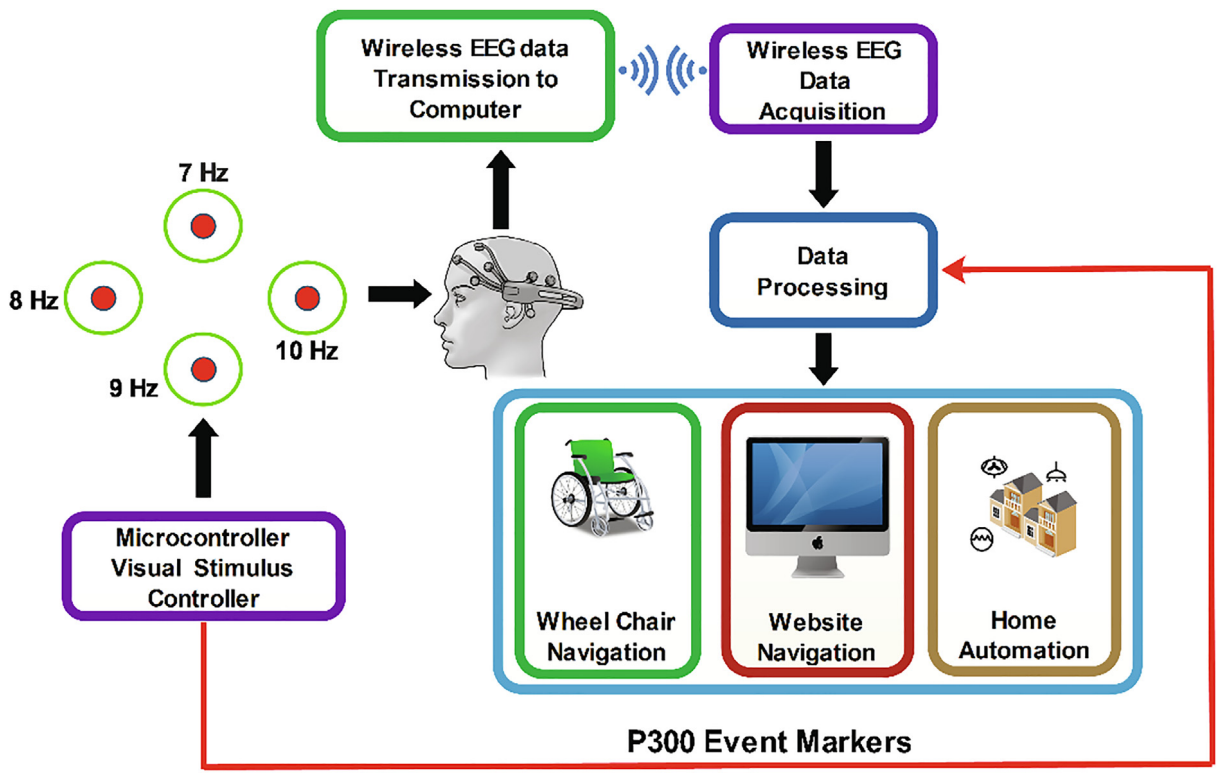

**Fig. 1.1.** Hybrid BCI utilising SSVEP and P300.

Each flash frequency for SSVEP that was generated precisely had a duty-cycle of 85% as that duty-cycle gave the highest performance as shown in a previous study [16]. The LED stimulus was driven using high-power MOSFETs (A09T) through a switch-down regulator, MP1584 to provide a constant current source of 3A, to provide optimum brightness throughout the experiment. The design of the regulator is shown in Fig. 1.2. The entire hardware was powered using a 12 V 10A battery source to avoid any mains interference. The complete schematic for the SSVEP radial stimulus is shown in Fig. 1.3. Four modules of the same design were used in this study with different flicker frequencies programmed for classification. The firm-ware was developed for producing the flicker, and the flicker accuracy was verified using a digital oscilloscope. The four frequency wave forms are shown in Figs. 1.4 and 1.5. The stimulus position and layout are shown in Fig. 1.6.

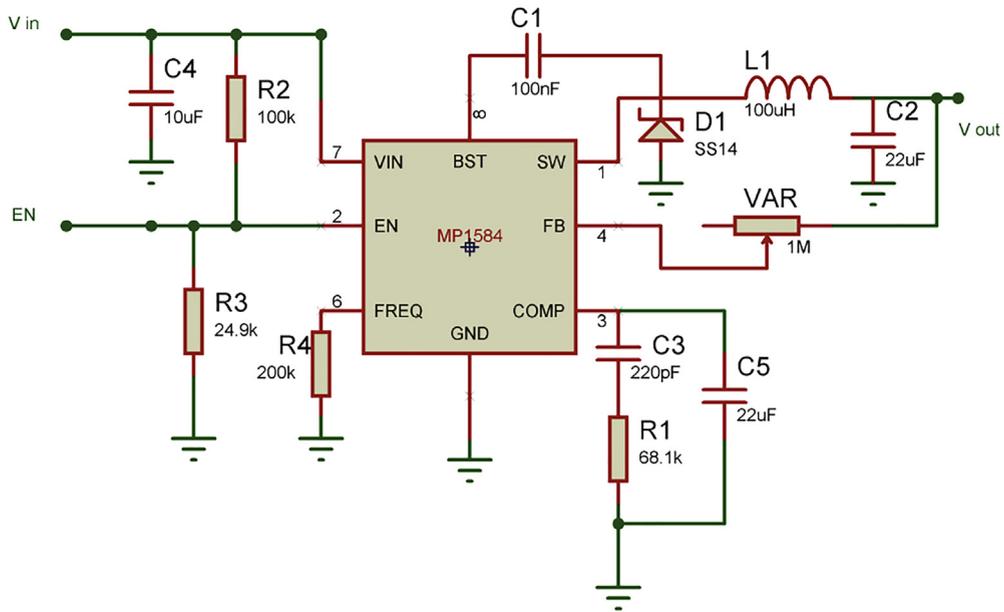

**Fig. 1.2.** High-speed switch down regulator for visual stimulus.

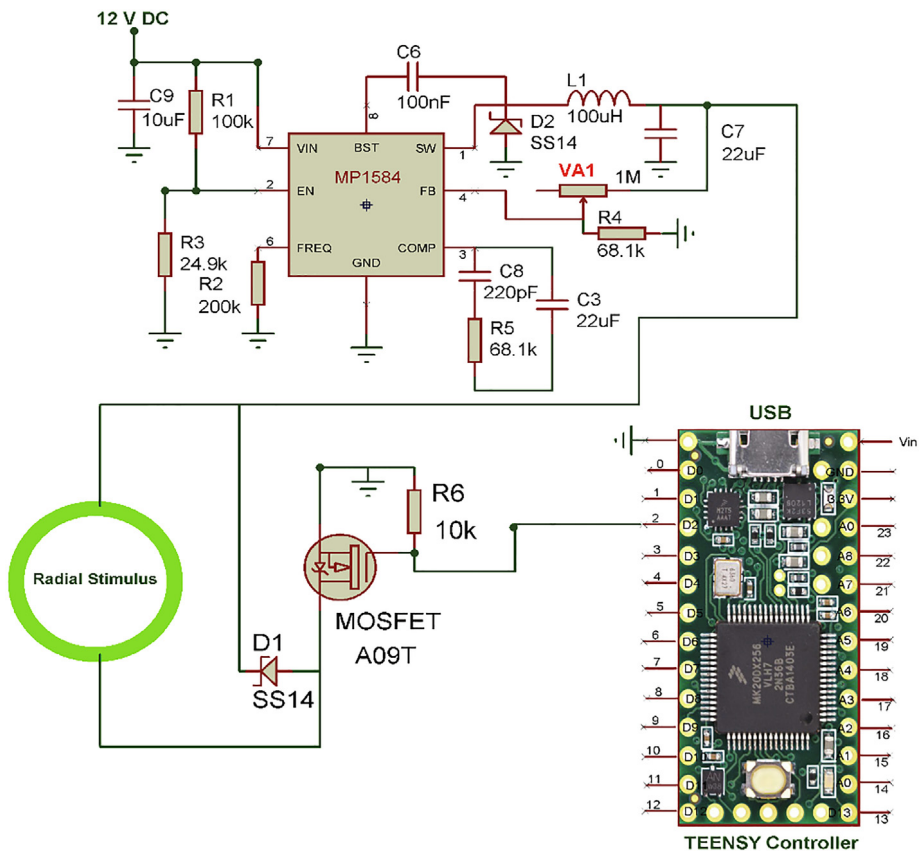

**Fig. 1.3.** SSVEP radial stimulus design.

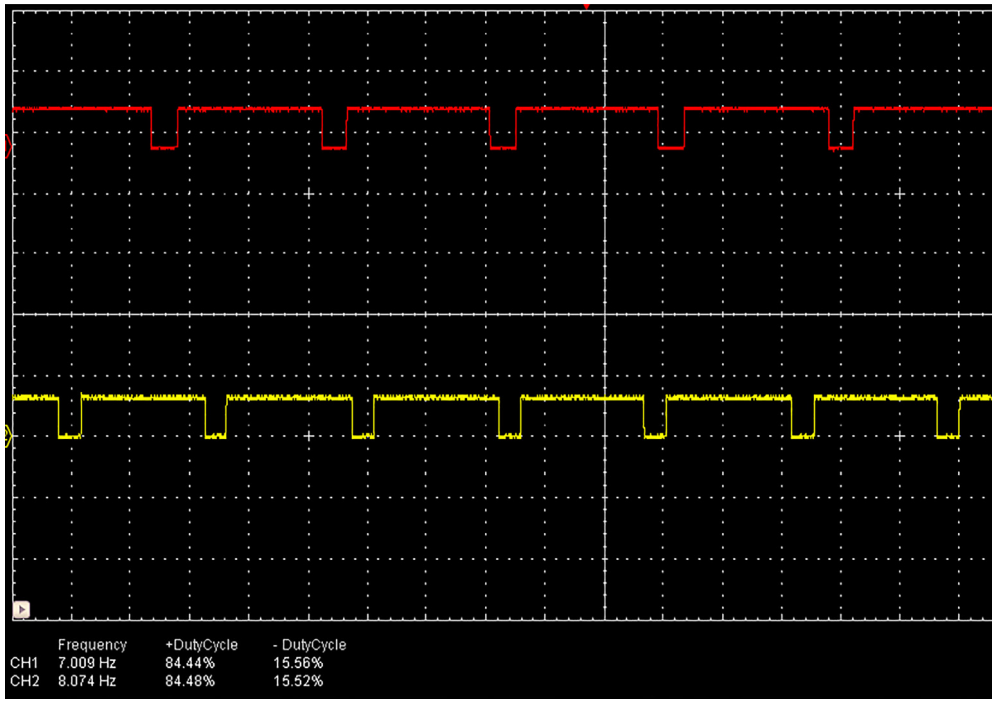

**Fig. 1.4.** 7 Hz and 8 Hz stimuli frequencies at 85% duty-cycle.

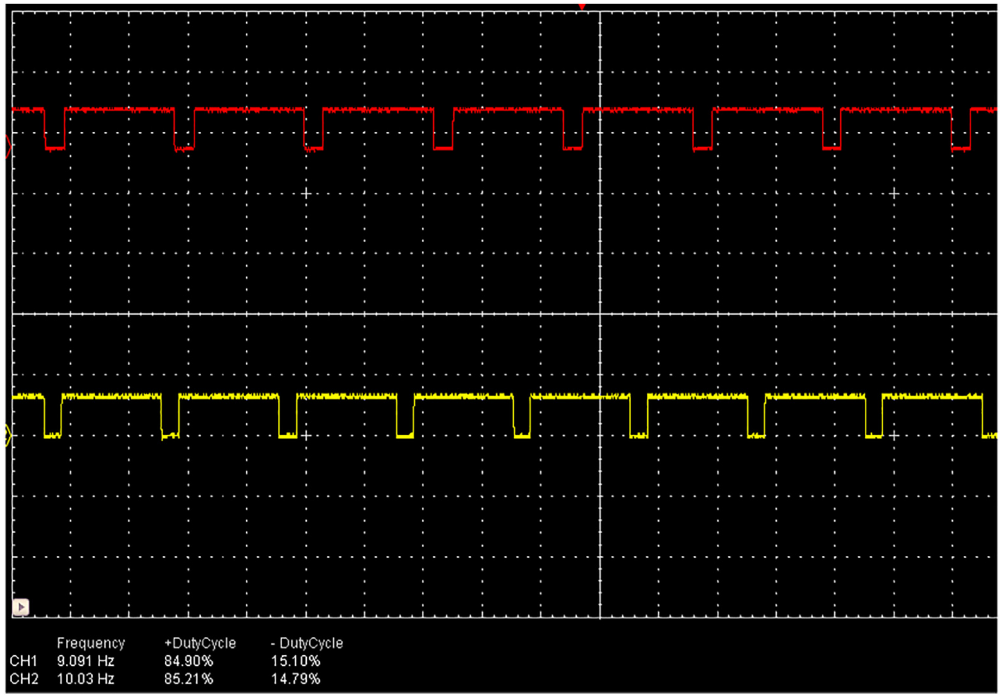

**Fig. 1.5.** 9 Hz and 10 Hz stimuli frequencies at 85% duty-cycle.

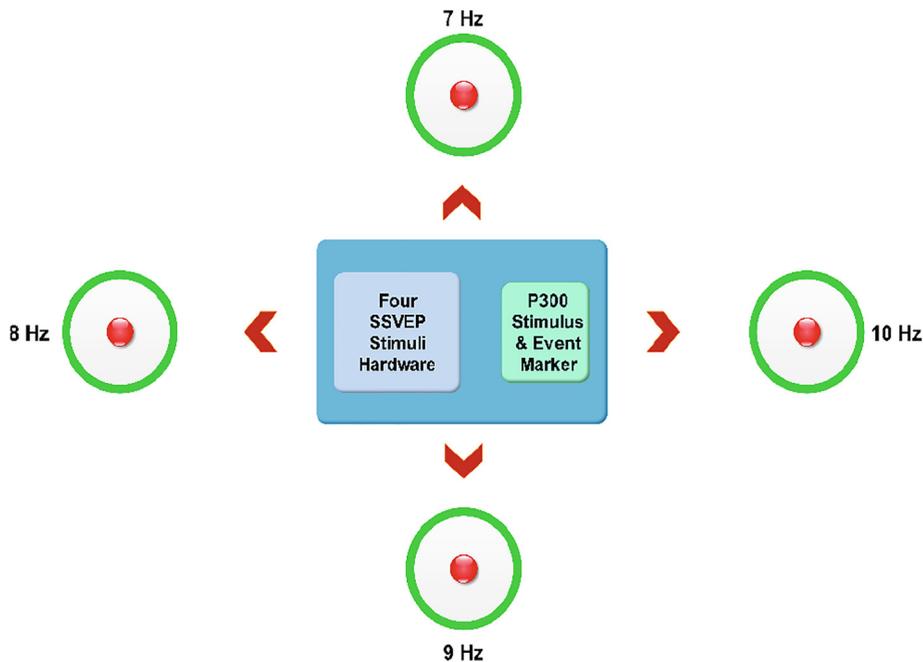

**Fig. 1.6.** Hybrid stimulus LED placement.

To evoke P300 component, four random flashes were generated using red LEDs and the flash event time markers were sent separately to the data recording software. The event markers from the microcontroller were then transferred using serial communication (Rx and Tx) to the computer. Microcontroller TTL levels were converted to RS232 using MAX3232. The red LED driver circuit was designed using high-current as mentioned previously. The complete schematic for the P300 flasher and event marker is shown in Fig. 1.7.

Randomly flashing four red LED timings were sent to the time marker values to the EMOTIV test bench software. For the LEDs located inside the radial rings, 7, 8, 9 and 10 Hz were marked as 111, 112, 113, 114, respectively and these values are stored together as a separate channel in the EEG data while recording. Fig. 1.8 shows the Testbench software with EEG data and the marker events. The random flash timings were set between 200 and 800 ms. The serial communication baud rate value was set as 115,200 on both transmitter and receiver side.

## 3. Design files

The design files are mainly for the microcontroller firmware which must be programmed individually for generating the stimuli frequencies.

| Design file name | File type | Open source license | Online |
| --- | --- | --- | --- |
| SSVEP radial stimulus | PDF | CC BY 4.0 | https://osf.io/8bc5s/ |
| 300 Stimulus and event marker | PDF | CC BY 4.0 | https://osf.io/8bc5s/ |
| Stimuli prototype | PDF | CC BY 4.0 | https://osf.io/8bc5s/ |
| Four_stimuli_flicker video | MP4 | CC BY 4.0 | https://osf.io/8bc5s/ |
| Test Bench data | PDF | CC BY 4.0 | https://osf.io/8bc5s/ |
| LEGO navigation | MP4 | CC BY 4.0 | https://osf.io/8bc5s/ |
| 7 Hz flicker firmware | Zip | CC BY 4.0 | https://osf.io/8bc5s/ |
| 8 Hz flicker firmware | Zip | CC BY 4.0 | https://osf.io/8bc5s/ |
| 9 Hz flicker firmware | Zip | CC BY 4.0 | https://osf.io/8bc5s/ |
| 10 Hz flicker firmware | Zip | CC BY 4.0 | https://osf.io/8bc5s/ |
| P300 firmware | Zip | CC BY 4.0 | https://osf.io/8bc5s/ |
| DOI | | | https://doi.org/10.17605/OSF.IO/8BC5S |

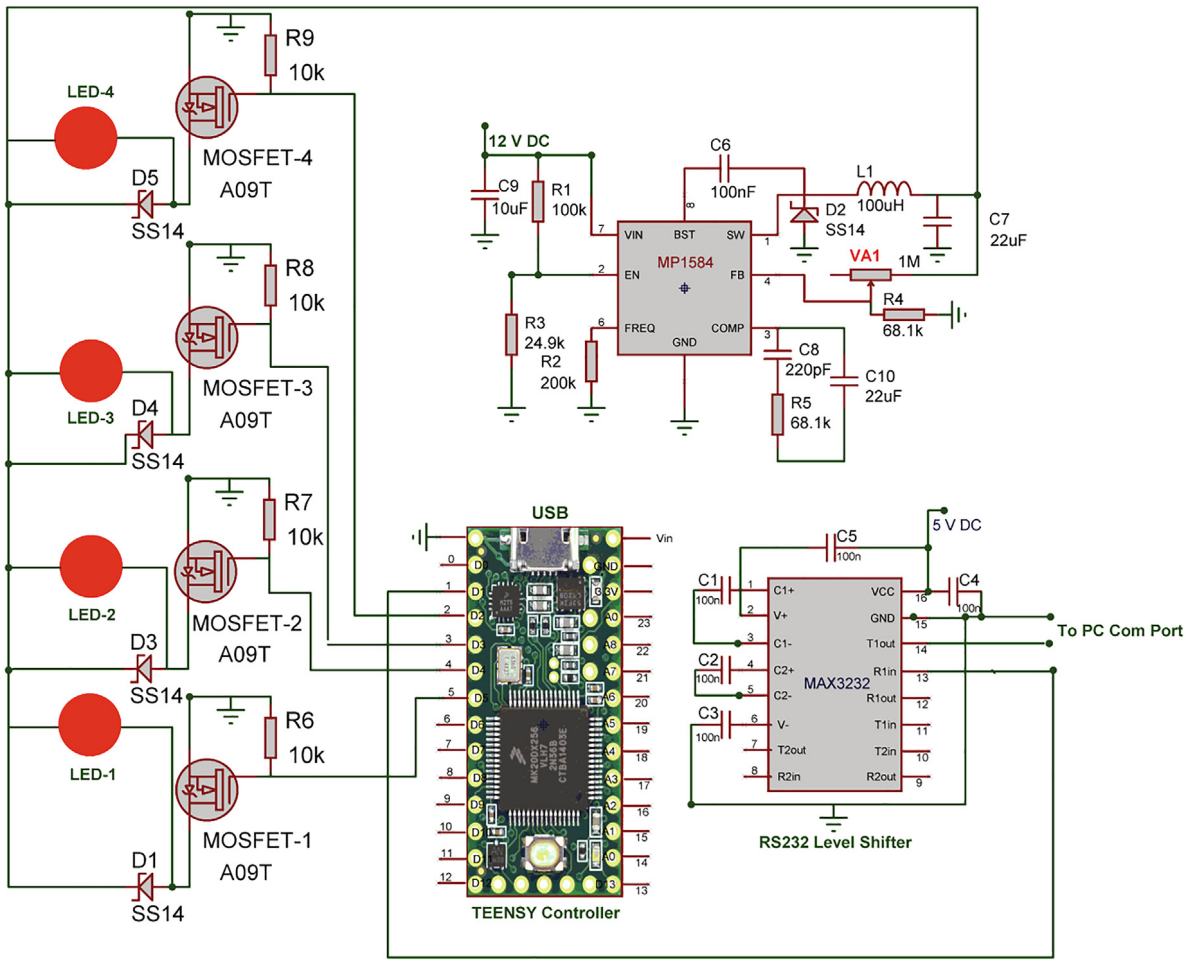

**Fig. 1.7.** P300 Stimulus and event marker.

## 4. Bill of materials

| Component | Quantity | Cost per unit (£) | Source of materials | Component description |
|---|---|---|---|---|
| Teensy 3.2 | 5 | 20.00 | Order Code: SC13983 http://cpc.farnell.com/teensy/teensy3-2/32-bit-arduino-comp-microcontroller/dp/SC13983 | Microcontroller Development Platform |
| COB radial LED ring | 2 | 19.00 | https://www.amazon.co.uk/Grandview-Multi-Color-Android-Bluetooth-headlight/dp/B072MRWB26/ref = sr_1_1?ie = UTF8&qid = 1507835174&sr = 8–1&keywords = COB + LED + Angel + Eyes + Ring | COB LED ring |
| 1 W Red LED | 4 | 0.15 | Order Code: LP379UHR1-C0G-01–00001 http://cpc.farnell.com/cree/lp379uhr1-c0g-01–00001/led-p4-red/dp/SC08569 | High power red LED |
| A3 Black Matte Acrylic board | 1 | 13.00 | https://www.ebay.co.uk/itm/Acrylic-Perspex-Colour-Clear-Mirror-Tinted-Frosted-Sheet-Cut-to-Size-Panel/201300425193?hash = item2ede70b9e9:m:mIRz75mK_Rd09f5zjHj1gmA | 5 mm A3 Perspex Sheet |
| MOSFETA09T | 8 | 0.27 | Order Code: 942-IRLML6344TRPBF https:// | N-channel 5A |

(continued)

| Component | Quantity | Cost per unit (£) | Source of materials | Component description |
|---|---|---|---|---|
| | | | www.mouser.co.uk/ProductDetail/Infineon-IR/IRLML6344TRPBF?qs=%2fha2pyFadugNaNXm2Hd8BxdVMKGEZ5IgToIu2OzAfARTwZF3dfzqp%252bnNlcjBcNvk | MOSFET |
| Schottky Diode1 N5819 | 8 | 0.28 | Order Code : 511–1 N5819 https://www.mouser.co.uk/ProductDetail/STMicroelectronics/1N5819?qs = sGAEpiMZZMvAvBNgSS9Lqo6cTfE%2fG8PN | Schottky Diode |
| MAX3232TTL/RS232 module | 1 | 3.00 | https://www.amazon.co.uk/MAX3232-Converter-Adaptor-Performance-Portable-Color-black/dp/B07SL14B5D/ | TTL to RS232Level Converter |
| P1584Regulator Module | 5 | 2.20 | https://www.hotmcu.com/mp1584-buck-step-down-3a-adjustable-regulator-module-p-82.html | Switching regulator |
| PCB | 1 | 7.99 | https://www.amazon.co.uk/DollaTek-Universal-Prototyping-Breadboard-Soldering/dp/B07DK52YK5/ | PCB |
| IC Socket | 1 | 4.99 | https://www.amazon.co.uk/sourcing-map-2–5-4 mm-Soldering-Adaptor/dp/B07H3SQL5K/ | IC Socket |
| Single Strandwire | 1 | 5.99 | https://www.amazon.co.uk/Hobby-Components-Ltd-22MM-SINGLE/dp/B01HZTLVBU/ | Wire |
| Double-sided tape | 1 | 5.99 | https://www.amazon.co.uk/MMOBIEL-Adhesive-Sticker-Digitizer-Smartphones/dp/B00PQO3HRS | Tape |
| Assorted Capacitors and Resistors pack | 1 | 5.00 | Online purchase | Assorted passive components |
| DOI | | | https://doi.org/10.17605/OSF.IO/8BC5S | |

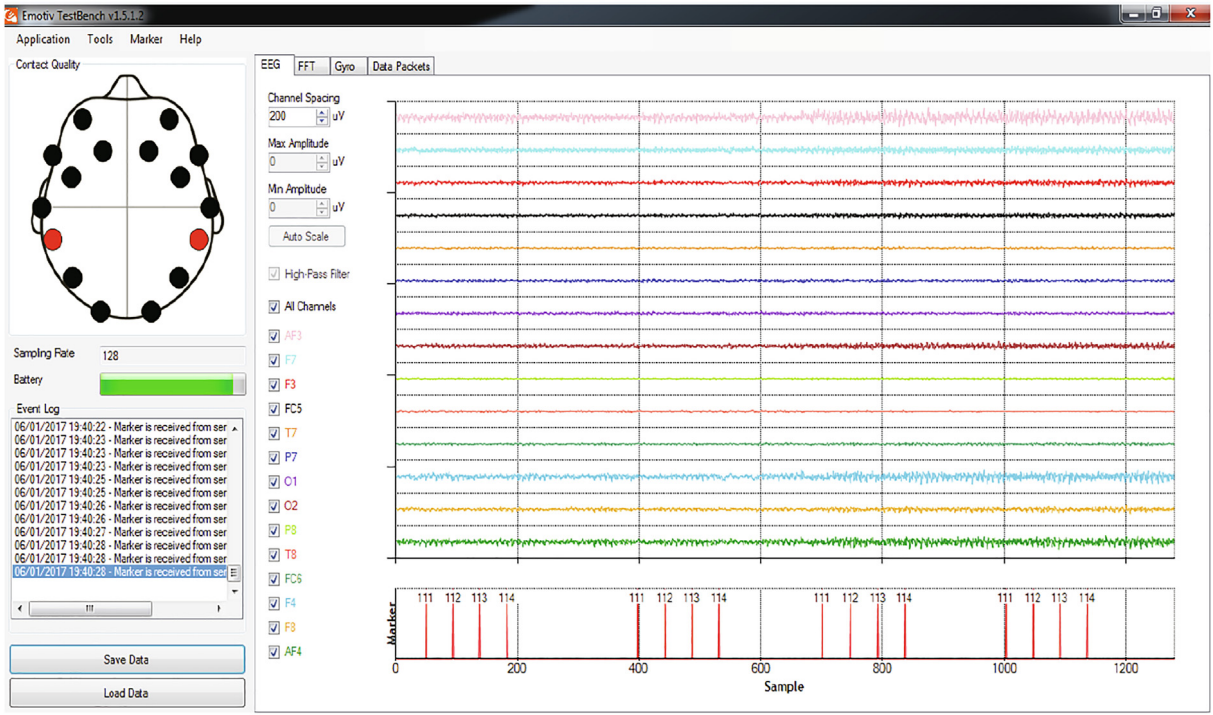

**Fig. 1.8.** Testbench software and event marker.

## 5. Build instructions

### 5.1. Hardware assembly

The system consists of two stimuli designs, one for SSVEP elicitation and the other for P300. For SSVEP, there were four individual modules for generating four different frequencies 7, 8, 9, and 10 Hz precisely. Each Teensy module was programmed with the developed firmware for the required flicker frequency, and the output from the MP1584 regulator was set to approximately 10.6 V DC and wired to the radial stimulus as in Fig. 1.3. Similarly, other three Teensy modules were programmed and wired accordingly.

For P300, the red LEDs were wired to the four output pins of the Teensy module as in Fig. 1.7 and the LEDs were placed at the centre of each COB ring. The MP1584 output needs to be set at 2.8 V DC for the optimum brightness for the red LEDs. For the serial communication, pin 1 of the Teensy module (Tx) needs to be wired to MAX3232 pin 13, which is serial data receive Rx.

### 5.2. Programming

The Teensy modules can be programmed directly through the USB port using the open-source Arduino IDE. The IDE installer can be downloaded from the Internet (www.arduino.cc) and needs to be installed for loading the firmware. Each Teensy module needs to be programmed individually with the required firmware for the selected stimulus location. The prototype was programmed with 7, 8, 9 and 10 Hz starting from the top COB LED stimulus in counter-clockwise direction.

### 5.3. Prototype

The prototype is shown in Fig. 1.9. The base board was made of Perspex with a black matte finish to avoid the light reflection with an A3 size (21 cm x29.7 cm). The P300 LEDs were fixed at the centre of the circular rings using double-sided glue tapes and wired to the hardware at the back of the board. The radial COB rings were soldered with thin flexible wires for connecting the MOSFET driver. The rings were fixed on the locations as shown in Fig. 1.9 using double-sided glue tapes. The control hardware was built on a general-purpose PCB with five 28 pin IC sockets and components wired using single-strand wires as per the schematic in Figs. 1.3 and 1.7.

## 6. Operation instruction

The prototype can be powered externally by a 12 V DC battery pack to avoid any external power interferences. P300 event marker cable needs to be attached to the serial port or can be used with a USB/serial converter cable. The baud-rate should be set to be the same value for both receiver and transmitter. Once powered, all the stimuli will be activated with the programmed stimulus frequency. Emotiv Test Bench application should display four serial values from the stimulus hardware corresponding to each visual stimulus. The required channels can be selected in the application to record the EEG data.

## 7. Validation and characterisation

The validation has been performed by investigating the classification accuracies of the hybrid BCI using SSVEP and P300. Five participants with perfect or corrected vision were chosen in the age group of 23 to 46 (three males and two females, mean age 35.6). The four radial stimuli along with four red LEDs were fixed on a black matte acrylic board at four locations as shown in the prototype (Fig. 1.9). The acrylic board was fixed at an eye level at 60 cm from each participant on an adjustable stand. For the EEG data collection, EMOTIV EPOC + research edition was used, which have 14 electrodes. For this study, electrode O2 was used for SSVEP detection as SSVEP responses are higher in the occipital region and F4 for the P300 responses, since that is the closest electrode to the midline for EMOTIV (EMOTIV has fixed electrode positions) where P300 responses are higher. The EMOTIV headset was prepared with all electrodes fitted with saline soaked felts and positioned on participants' head. The connection quality of the electrodes was tested using the test bench software, and further saline was added as necessary.

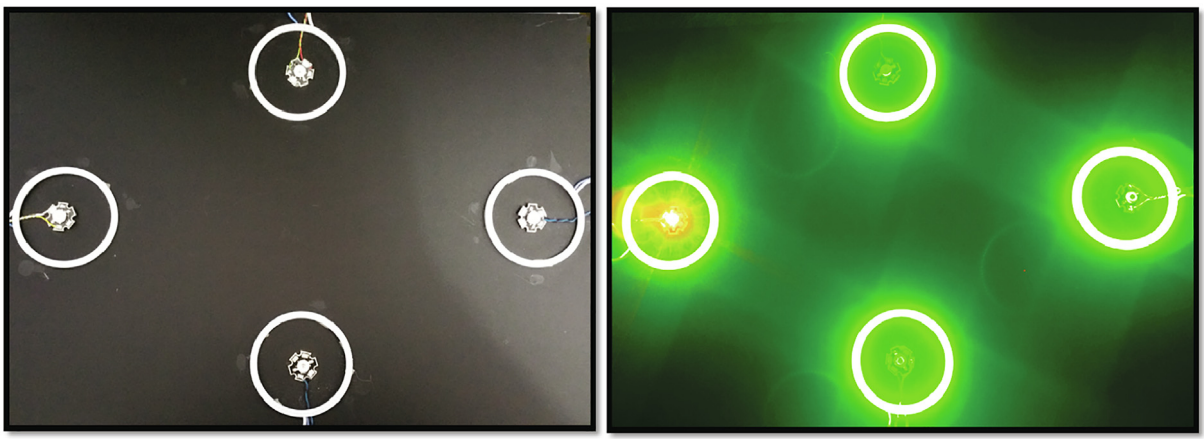

**Fig. 1.9.** Prototype.

The EEG recording process started with the 7 Hz stimulus for three seconds followed by five-second rest period when the participant looked away from the flashing stimulus. This was followed by stimuli with frequencies of 8, 9 and 10 Hz with the same rest period to finish one complete session. Five sessions were recorded for each participant, which had EEG data for both SSVEP and P300 events. The stimulus timings are shown in Fig. 1.10 for each frequency and rest period. Three seconds on the timing chart is the focus time when the participants pay attention to each stimulus, and five seconds denotes the rest time between the changeovers from one stimulus to the next one in a sequential order. The recorded EEG data in EDF format was converted to MATLAB format using EEGLAB. Codes were then developed in MATLAB to process data from both SSVEP and P300 event detections for evaluating the hybrid stimuli. Fig. 1.11 shows a photo of Lego control using this hybrid BCI.

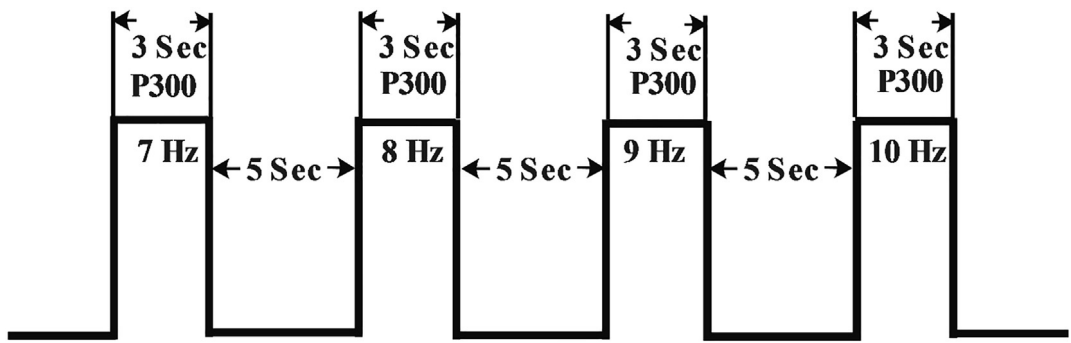

**Fig 1.10.** Stimulus timing.

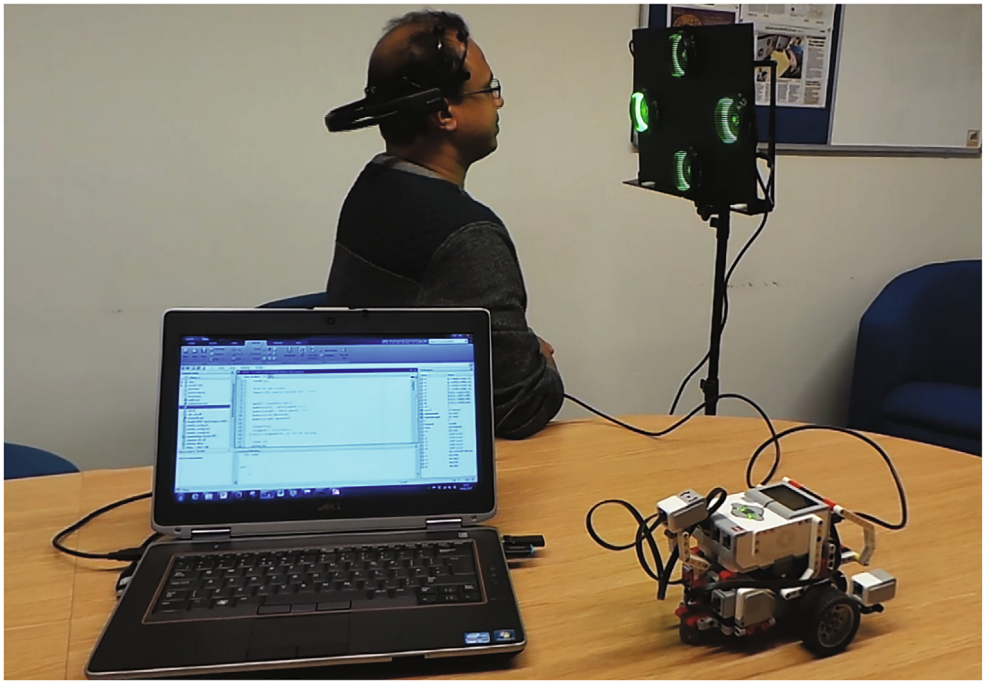

**Fig 1.11.** Lego control using hybrid BCI.

## 7.1. Data analysis

For SSVEP analysis, the data was extracted from the channel O2 for the required three-second time period for each frequency and stored for further processing. For SSVEP data classification, the principal frequencies along with first and second harmonics were analysed. The stored EEG data was filtered with a band-pass filter of 2 Hz bandwidth with centre frequency as the stimulus frequency. This was performed for all the principal frequencies and their harmonics. The variance of each filtered data was computed (this represented the energy of the signal), which included the principal frequency and with the two harmonics added and stored for classification analysis.

For the P300 analysis, the data was extracted from channel F4 along with the four event markers. The event markers were set as 111 for 7 Hz, 222 for 8 Hz, 333 for 9 Hz and 444 for 10 Hz. For the same time window as in SSVEP analysis as before when the participant focused on a stimulus, the algorithm checked for the event markers to find the peak value within 600 ms from the event marker. This was performed for all four markers for four frequencies. The peak values for all the frequencies for the required time window was stored to analyse the classification accuracy.

## 7.2. Results

The study explored the possibilities of combining two EEG paradigms to develop a hybrid BCI visual stimuli hardware with reduced visual stimulus focus time and to allow good single trial classification output. The developed standalone hybrid stimuli successfully generated frequencies 7, 8, 9 and 10 Hz that had a narrow frequency gaps between them. P300 events were also generated simultaneously with four event markers and were successfully detected in the recorded EEG using MATLAB. From the analysis of data from the five participants, both SSVEP and BCI yielded better classification rates individually for the same time window for each frequency. P300 also gave good performance, which is unusual for single trial analysis. This combined classification results from SSVEP and P300 improved the reliability in classification for controlling external application within a short time window of three seconds.

Moreover, SSVEP based command control was successfully implemented using the hybrid visual stimuli to control the movements of a LEGO robot (this could be used to control a wheelchair by the severely paralysed in the real world). The visual stimuli with four low-frequency flicker was presented for the user to focus one at a time to perform the motion commands to the LEGO robot. The commands were sent to the robot with a minimum attention time of 3 – 4 s for each visual stimulus. The movements of the robot were near real-time apart from minor delays in data processing and classification. The minor delays were approximately 3–4 sec for each motion, and this was due to the data processing time required before the action.

The hardware platform could be used in various neurological investigations or psychological studies as a stand-alone visual stimulus device or with event marker-related experiments using P300 events. The device could be easily customised for single or multiple stimuli generations as required.

## 8. Conclusion

We presented a low-cost hybrid visual stimulus design and implementation using readily available hardware, which could be used for practical BCI based applications as well as for SSVEP and P300 research studies (DOI:https://doi.org/10.17605/OSF.IO/8BC5S). The hardware design was discussed in detail, and the design files can be downloaded from Open Science Framework (https://osf.io/8bc5s/). The hardware can be built easily as the main components are available as a module which needs to be soldered on a general-purpose PCB or also can be constructed on a breadboard. Since the open-source approach was followed, the design can be reused, altered or adapted for similar projects for BCI studies or other research purposes.

**Declaration of Competing Interest**

The authors declare that they have no known competing financial interests or personal relationships that could have appeared to influence the work reported in this paper.

**Acknowledgements**

This research did not receive any specific grant from funding agencies in the public, commercial, or not-for-profit sectors.

**Human rights**

All the experiments carried out with human participants were in accordance with The Code of Ethics of the World Medical Association (Declaration of Helsinki). The informed consents were obtained from the participants for experiments. All the collected data were anonymised to maintain the privacy of participants. The experiment also received ethical approval from the Faculty of Sciences Ethics Committee at University of Kent.

## References

[1] D.J. Mcfarland, J.R. Wolpaw, Brain–Computer Interfaces for the Operation of Robotic and Prosthetic Devices, (2010) 169–187. https://doi.org/10.1016/s0065-2458(10)79004-5.
[2] U. Chaudhary, N. Birbaumer, M.R. Curado, Brain-Machine Interface (BMI) in paralysis, Ann. Phys. Rehab. Med. 58 (2015) 9–13, https://doi.org/10.1016/j.rehab.2014.11.002.
[3] S.R. Soekadar, N. Birbaumer, M.W. Slutzky, L.G. Cohen, Brain–machine interfaces in neurorehabilitation of stroke, Neurobiol. Dis. 83 (2015) 172–179, https://doi.org/10.1016/j.nbd.2014.11.025.
[4] S. Mouli, R. Palaniappan, I.P. Sillitoe, J.Q. Gan, Performance analysis of multi-frequency SSVEP-BCI using clear and frosted colour LED stimuli, in: 13th IEEE International Conference on BioInformatics and BioEngineering, 2013, pp. 1–4, https://doi.org/10.1109/BIBE.2013.6701552.
[5] S. Mouli, R. Palaniappan, I.P. Sillitoe, A configurable, inexpensive, portable, multi-channel, multi-frequency, multi-chromatic RGB LED system for SSVEP Stimulation, in: A.E. Hassanien, A.T. Azar (Eds.), Brain-Computer Interfaces: Current Trends and Applications, Springer International Publishing, 2015, pp. 241–269, https://doi.org/10.1007/978-3-319-10978-7_9.
[6] M. Spüler, A Brain-Computer Interface (BCI) system to use arbitrary Windows applications by directly controlling mouse and keyboard, in 37th Annual International Conference of the IEEE Engineering in Medicine and Biology Society (EMBC), (2015) 1087-1090. DOI:10.1109/EMBC.2015.7318554.
[7] Y. Liu, X. Jiang, T. Cao, F. Wan, P.U. Mak, P. Mak, M.I. Vai, Implementation of SSVEP based BCI with Emotiv EPOC, in: Proceedings International Conference on Virtual Environments Human-Computer Interfaces and Measurement Systems (VECIMS), 2012, pp. 34–37, https://doi.org/10.1109/vecims.2012.6273184.
[8] M.H. Chang, J.S. Lee, J. Heo, K.S. Park, Eliciting dual-frequency SSVEP using a hybrid SSVEP-P300 BCI, J. Neurosci. Methods 258 (2016) 104–113, https://doi.org/10.1016/j.jneumeth.2015.11.001.
[9] A. Kubacki, A. Jakubowski, Controlling the industrial robot model with the hybrid BCI based on EOG and eye tracking, AIP Conf. Proc. 2029 (2018), 020032.
[10] Z. Wang, Y. Yu, M. Xu, Y. Liu, E. Yin, Z. Zhou, Towards a hybrid BCI gaming paradigm based on motor imagery and SSVEP, Int. J. Hum.-Comput. Interaction 35 (2019) 197–205, https://doi.org/10.1080/10447318.2018.1445068.
[11] D. Kapgate, D. Kalbande, U. Shrawankar, An optimized facial stimuli paradigm for hybrid SSVEP+P300 brain computer interface, Cogn. Syst. Res. 59 (2020) 114–122, https://doi.org/10.1016/j.cogsys.2019.09.014.
[12] I. Choi, I. Rhiu, Y. Lee, M.H. Yun, C.S. Nam, A systematic review of hybrid brain-computer interfaces: taxonomy and usability perspectives, PLoS One 12 (2017) e0176674, https://doi.org/10.1371/journal.pone.0176674.
[13] A. Combaz, M.M. Van Hulle, Simultaneous detection of P300 and steady-state visually evoked potentials for hybrid brain-computer interface, PLoS One 10 (2015) e0121481, https://doi.org/10.1371/journal.pone.0121481.
[14] S. Amiri, A. Rabbi, L. Azinfar, R. Fazel-Rezai, A review of P300, SSVEP, and hybrid P300/SSVEP, Brain- Comput. Interface Syst. (2013), https://doi.org/10.5772/56135.
[15] M. Wang, I. Daly, B.Z. Allison, J. Jin, Y. Zhang, L. Chen, X. Wang, A new hybrid BCI paradigm based on P300 and SSVEP, J. Neurosci. Methods 244 (2015) 16–25, https://doi.org/10.1016/j.jneumeth.2014.06.003.
[16] S. Mouli, R. Palaniappan, Toward a reliable PWM-based light-emitting diode visual stimulus for improved SSVEP response with minimal visual fatigue, J. Eng. 2 (2017) 7–12, https://doi.org/10.1049/joe.2016.0314.